\documentclass{amsart}
\usepackage{amsmath,amssymb,amscd,bbm,dsfont,enumerate}
\usepackage{psfrag}
\usepackage{graphicx}
\usepackage[round]{natbib}

\usepackage{mathtools}

\newcommand{\R}{\mathbb R}  
 
\renewcommand{\P}{\mathbbm P} 
\newcommand{\E}{\mathbb E} 
\newcommand{\cf}{\mathbbm 1} 
\newcommand{\OM}{\mathcal{A}_{\text{OM}}} 
\newcommand{\OMF}{\mathcal{F}_{\text{OM}}}

\newcommand{\beq}[1]{\begin{equation}\label{#1}}
\newcommand{\eeq}{\end{equation}}
\newcommand{\beqn}[1]{\begin{equation}\nonumber}
\newcommand{\vol}{\text{vol}}
\newcommand{\dd}{\mathrm d}

\title[The correct cost functional for variational assimilation]{What is the correct cost functional for variational data assimilation?}
\author{Jochen Br\"{o}cker}
\address{School of Mathematical and Physical Sciences, University of Reading, United Kingdom, \today}
\thanks{Published in {\em Climate Dynamics} (2018), doi:~10.1007/s00382-018-4146-y.
The author was supported by the UK Engineering and Physical Sciences Research Council under grant agreement EP/L012669/1.
Fruitful discussions with Tobias Kuna, Dan Crisan, Andrew Stuart, Colin Cotter, St\'{e}ph\`{a}ne Vannitsem and Horatio Boedihardjo are gratefully acknowledged.
}
\begin{document}
\maketitle  
% 
% \linenumbers
% 
\begin{abstract}	
Variational approaches to data assimilation, and weakly constrained four dimensional variation (WC-4DVar) in particular, are important in the geosciences but also in other communities (often under different names).
The cost functions and the resulting optimal trajectories may have a probabilistic interpretation, for instance by linking data assimilation with Maximum Aposteriori (MAP) estimation. 
This is possible in particular if the unknown trajectory is modelled as the solution of a stochastic differential equation (SDE), as is increasingly the case in weather forecasting and climate modelling.
In this case, the MAP estimator (or ``most probable path'' of the SDE) is obtained by minimising the Onsager--Machlup functional.
Although this fact is well known, there seems to be some confusion in the literature, with the energy (or ``least squares'') functional sometimes been claimed to yield the most probable path.
The first aim of this paper is to address this confusion and show that the energy functional does not, in general, provide the most probable path. 
The second aim is to discuss the implications in practice.
Although the mentioned results pertain to stochastic models in continuous time, they do have consequences in practice where SDE's are approximated by discrete time schemes. 
It turns out that using an approximation to the SDE and calculating its most probable path does not necessarily yield a good approximation to the most probable path of the SDE proper.
This suggest that even in discrete time, a version of the Onsager--Machlup functional should be used, rather than the energy functional, at least if the solution is to be interpreted as a MAP estimator.
\end{abstract}
\section{Introduction}
% 
%%%%%%%%%%%%%%%%%%%%%%%%%%%% 
% 
In the geosciences, the term data assimilation refers to a variety of mathematical and numerical techniques whereby time series of observations are employed to estimate states or trajectories of relevant dynamical models.
In other words, plausible states or orbits are determined which, on the one hand, are consistent with a given dynamical model and, on the other hand, are consistent with a given set of observations.
Many different approaches to data assimilation exist, based on very different philosophies and premises, see for instance~\cite{ide97,kalnay01,evensen_enkf_2007}, but this list is by no means complete.
Both within the atmospheric sciences, but also in other branches of physics and engineering, variational approaches have gained widespread attention (although the nomenclature may differ considerably).
A particular instance of this idea is known as weakly constrained four dimensional variation (WC-4DVar) in atmospheric sciences; basically, a series of model states is found that minimises a cost functional which quantifies both the deviations from the observed data as well as the misfit with the given model. 
An early paper on discrete time WC-4DVar in atmospheric sciences is~\cite{derber89}, see also~\cite{kalnay01}.
The cost function is almost invariably some form of quadratic error, and for this reason, the technique is known as the minimum energy estimator in the engineering community, see for instance~\cite{JAZ} or~\cite{mortensen_likelihood_filtering_1968} (in the latter publication, the authors go further and derive an incremental version).
In the atmospheric sciences and in particular in climate modelling, stochastic models are becoming ever more important, despite having a long and distinguished history already~\citep[see for instance][and references therein]{imkeller_stochastic_climate_models_1999,franzke_stochastic_climate_2014}.
Mathematically speaking, climate models increasingly take the form of stochastic differential equations (SDE's).
Consequently, data assimilation into such models needs well understood foundations.
In particular, if variational data assimilation into SDE's is envisaged, the question arises as to what cost function should be used, and in particular whether the cost functions and the resulting optimal trajectories have any probabilistic interpretation.
A possible avenue is to link variational data assimilation with Maximum Aposteriori (MAP) estimation. 
The MAP estimator of a random variable given some observations is essentially the maximiser of the posterior, that is, of the conditional density of the unknown random variable given the observations.
In some sense, the MAP estimator can be interpreted as the ``most probable value'' of the unknown random variable given the observation.
The concept of density generalises to situations where the unknown random variable is an entire function, given by the solution of a stochastic differential equation (SDE), and the MAP estimator becomes the ``most probable path'' of the SDE (see e.g.~\cite{zeitouni_maximum_1987},~\cite{zeitouni_existence_1988}; for MAP estimation in classical inverse problems but with random observations see~\cite{cotter_bayesian_inverse_2009}; see also~\citet{apte_sampling_posterior_2007,stuart_inverse_bayesian_2010} for applications to Bayesian estimation in stochastic dynamical systems).
Contrary to what is sometimes claimed in the literature, the most probable path of an SDE is not a minimiser of the energy functional but rather of the Onsager--Machlup functional, which differs from the energy functional in that the latter contains extra terms. 
In other words, to find MAP estimators or most probable paths for SDE's, the Onsager--Machlup functional has to be minimised, rather than the energy functional.
The first aim of this paper is to illustrate this well known fact. 
The reader is referred to~\cite{zeitouni_maximum_1987},~\cite{zeitouni_existence_1988} for a rigorous derivation of the Onsager--Machlup functional and discussion of the MAP estimator in the context of SDE's.
The second aim is to show that although this is a result pertaining to stochastic models in continuous time, it does have consequences in discrete time. 
In practice, SDE's are approximated by discrete time schemes, for instance the Euler scheme which results in discrete time stochastic dynamical system with additive Gaussian errors. 
The (negative logarithm of the) density of solutions to this discrete time system is given by the energy functional. 
But we will argue that the appropriate functional in this situation should {\em still} be the Onsager--Machlup functional or a discrete time version thereof, at least if the solution is to be interpreted as a MAP estimator.
The reason is that the MAP estimator (or most probable path) of an approximation to the SDE is not necessarily a good approximation to the most probable path of the SDE proper, as we will see.
It is worth noting that this point involves the dynamics only and is entirely independent of whether observations are considered discrete or continuous in time.
In Section~\ref{sec:map_definition}, we revisit the concepts of densities for random variables and the MAP~estimator.
In Section~\ref{sec:map_estimators_sde}, we specialise to the situation were the unknown random variable is a trajectory of a stochastic differential equation, and demonstrate that the energy functional cannot be the correct functional to determine the MAP~estimator.
An expression for the Onsager--Machlup functional will also be provided.
The findings will be supported by numerical simulations in Section~\ref{sec:numerical_experiment}.
Further, these simulations illustrate that the Onsager--Machlup functional essentially provides the correct density for paths of SDE's even though the simulations are not truly continuous in time but rather use an approximation scheme that is discrete in time.
Section~\ref{sec:auxilliary} provides the Onsager--Machlup functional for more general SDE's that are not used in the present paper but which are relevant for the climate sciences, namely SDE's with multiplicative noise~\citep[see e.g.][]{franzke_stochastic_climate_2014}\footnote{We are grateful to referee St\'{e}ph\`{a}ne Vannitsem for stressing this point.}.
Section~\ref{sec:conclusions} concludes with a discussion as to how our findings bear on discrete time simulations of SDE's. 
An informal derivation of the Onsager--Machlup functional is provided in Appendix~\ref{sec:correct_functional}. 
\section{Definition of the Maximum Aposteriori (MAP) estimator}
\label{sec:map_definition}
A fundamental concept in statistics in general and data assimilation in particular is the Maximum Aposteriori or MAP estimator. 
Let $X, Y$ be random variables, where we interprete $X$ as the unknown quantity (to be estimated) and $Y$ as the observation.
Let $p(x|y)$ denote the conditional probability density function of $X$ given that $Y$ assumes the value $y$.
A {\em MAP~estimator} of $X$ given $Y$ is a maximiser over $x$ of the density $p(x|y)$.
That is, the MAP estimator is a function $\hat{x}(y)$ so that for any $y$ we have
\beqn{equ:mapestimatorI}
p(\hat{x}(y)|y) = \sup_{x} p(x|y).
\eeq
MAP estimators need not exist in general, nor are they unique.
Since the observations $Y$ play the role of parameters in this problem, they will mostly suppressed in the notation for the sake of simplicity.
That is, if $X$ is a random variable with density $p_{X}$, we understand that $p_{X}$ might in fact be the conditional density of $X$ given some observations or parameters.
The presented definition of the MAP estimator will be referred to as the {\em de facto} definition (following~\cite{abreu_map_discretisation_2014}); there is an alternative definition which not only provides an intuitive interpretation but is more generally applicable.
Roughly speaking, the MAP estimator of a random variable $X$ is the center of a small ball positioned so as to have greatest possible probability of containing $X$, in the limit of the diameter of that ball going to zero.
More formally, suppose that $X$ is a random variable with values in some vector space $V$ with norm $\|.\|$.
Then the MAP estimator is a point $\hat{x}$ so that for any other point $x$  
\beq{equ:mapestimatorII}
\limsup_{\epsilon \to 0} 
\frac{\P(\|X - x\| \leq \epsilon)}%
    {\P(\|X - \hat{x}\| \leq \epsilon)}
\leq 1.
\eeq
If observations are present, then these probabilities are conditional probabilities given those observations.
If a random variable $X$ with values in $\R^d$ has a density $p$ which is everywhere positive, then a MAP estimator according to the alternative definition~\eqref{equ:mapestimatorII} is a MAP estimator according to the {\em de facto} definition and vice versa.
Indeed, if $X$ has a positive density $p$, then for all $x \in \R^d$ the relation
\beq{equ:lebesguediff}
p(x) = \lim_{\epsilon \to 0} \frac{\P(\|X - x\| \leq \epsilon)}{\vol\{z \in \R^n; \|z\| \leq \epsilon\}}
\eeq
holds (except perhaps if $x$ is in some exceptional set which has however volume zero; we will ignore this technical point).
Here, $\vol$ denotes the standard volume on $\R^d$.
Hence, if $y$ is so that $p(y) > 0$, then for any $x$ we have 
\beq{equ:lebesguediffII}
\begin{split}
\frac{p(x)}{p(y)}
 & = \lim_{\epsilon \to 0} 
    \frac{\P(\|X - x\| \leq \epsilon)}%
    {\vol\{z \in \R^d; \|z\| \leq \epsilon\}} 
    \lim_{\epsilon \to 0} 
    \frac{\vol\{z \in \R^d; \|z\| \leq \epsilon\}}%
    {\P(\|X - y\| \leq \epsilon)}\\
 & = \lim_{\epsilon \to 0} 
    \frac{\P(\|X - x\| \leq \epsilon)}%
    {\P(\|X - y\| \leq \epsilon)}.
\end{split}
\eeq
The relation~\eqref{equ:lebesguediffII} shows that any point $\hat{x} \in \R^d$ which satisfies the de facto definition of a MAP estimator will also satisfy the alternative definition and vice versa. 
A strong point of the {\em de facto} definition is that it provides a means to find a MAP estimator through an optimisation problem. 
An important insight from the alternative definition though is that it is not quite necessary to have a probability density function as in Equation~\eqref{equ:lebesguediff} in order to define the MAP estimator. 
In particular the normalisation in Equation~\eqref{equ:lebesguediff} need not be the standard volume; normalising in a different way would give a different density, but as long as the normalisation is the same for all reference points $x$ and the resulting density is still everywhere positive, we would obtain the same MAP estimators, since the relation~\eqref{equ:lebesguediffII} would still be valid.
For instance, if $W$ is another random variable, we could normalise as follows
\beq{equ:densitydef}
p^{(W)}(x) = \lim_{\epsilon \to 0} \frac{\P(\|X - x\| \leq \epsilon)}{\P(\|W\| \leq \epsilon)}
\eeq
if the limit exists for every $x$; if $p^{(W)}$ is everywhere positive, $p^{(W)}$ can be used to calculate the MAP just as well.
It turns out that generalised densities as in Equation~\eqref{equ:densitydef} might still be well defined even if $X$ has values in some infinite dimensional space with norm $\|.\|$ for which there exists no generalisation of the standard volume.\footnote{%
The problem is the translation invariance of the standard volume.
In an infinite dimensional normed space, a ball of unit radius may contain infinitely many disjoint balls of sufficiently small but nonzero radius.
By translation invariance, these balls must have the same volume. % 
But this means that either the volume of the unit ball is infinity or the volume of a sufficiently small ball is zero.}
This is precisely the situation when trying to find MAP estimators for trajectories of continuous time stochastic dynamical models; such a trajectory is a function (of time) and hence an infinite dimensional object.
Hence the Definition~\eqref{equ:lebesguediffII} of a density does not apply in this situation but Definition~\eqref{equ:densitydef} does, provided we find a suitable random variable $W$ to normalise with.
\section{MAP estimators for stochastic difference and differential equations}
\label{sec:map_estimators_sde}
\begin{figure}
\begin{center}
\includegraphics{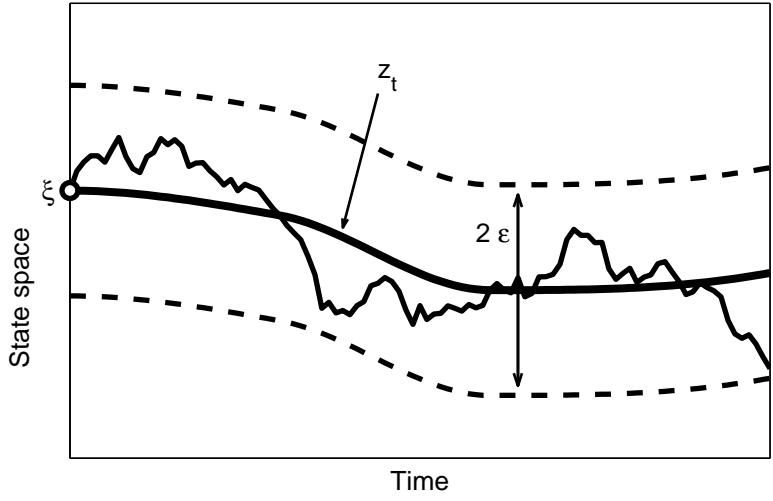}
\end{center}
\caption{\label{fig:sausages} The plot shows the event that the solution $\{X_t, t \in I\}$ of the SDE~\eqref{equ:continuousdynamics} (thin solid line) falls entirely into a small strip of width $\epsilon$ around the reference trajectory $\{z_t, t \in I\}$ (thick solid line). The strip is indicated with dashed lines.
(Note that this is a schematic sketch rather than an actual simulation.)
% new
}
\end{figure}

The link between MAP estimators and data assimilation in discrete time can be described as follows.
The dynamics underlying the observations is modelled as a stochastic difference equation of the form
\beq{equ:discretedynamics}
X_{n} = F(X_{n-1}) + R_n, \qquad n = 1, 2, \ldots, N
\eeq
where $F$ is some mapping on a vector space $E$ (called the state space), and the $R_n, n = 1, 2, \ldots$ are taken as independent and identically distributed random variables with values in $E$.
For simplicity's sake, we assume throughout that $E$ is one~dimensional (see however Sec.~\ref{sec:auxilliary}).
Further, the $R_n, n = 1, 2, \ldots$ are assumed to be normal with mean zero and variance $\gamma$.
We further set $X_0 = \xi$, where $\xi \in E$ is known. 
The observations are assumed to be functions of the $X_1, \ldots, X_n$ further corrupted by noise. 
But as said earlier, they will enter the densities as parameters in some way which is not relevant for our purposes.
It is then a simple matter to show that
\beq{equ:apostdensity}
\begin{split} 
\lim_{\epsilon \to 0} 
    \frac{\P(\max_n |X_n - x_n | \leq \epsilon)}%
{\P(\max_n |R_n| \leq \epsilon)}
& = \frac{p_{X_1, \ldots, X_N}(x_1, \ldots, x_N)}%
{p_{R_1, \ldots, R_N}(0, \ldots, 0)}\\
& = \exp \left( -\frac{1}{2\gamma}
    \sum_{n = 1}^{N} 
        (x_n - F(x_{n-1}))^2\right),
\end{split}
\eeq     
where we understand that $x_0 = \xi$.
Since $(X_1, \ldots, X_N)$ is a random variable in $E^N$, we can interprete the right hand side of Equation~\eqref{equ:apostdensity} as a density of $(X_1, \ldots, X_N)$ according to Definition~\eqref{equ:densitydef} with $V = E^N$ and norm $\|(x_1, \ldots, x_N)\| = \max_n |x_n|$.
Atmospheric and ocean dynamics are, however, continuous in time, as are many other processes in science and engineering where data assimilation is relevant.
Considering data assimilation in discrete time is merely a concession to practical constraints.
Indeed, there are several different processess that introduce time stepping in operational practice, for instance the integration of the model or the batch processing of the observations, but the relevant time steps can be very different.
Accounting for ``model error'' with additive noise after discretising models in time will result in the solutions for different time stepping having different statistical properties. 
Although this is to some extent inevitable, we still ought to have a formalism for comparing these different solutions, as they ultimately represent the same thing.
A convenient way to enable comparison of different discretisations (with noise added) is to formulate a stochastic model in continuous time, that is, a stochastic differential equation (SDE), and consider any discretisation as an approximation of that model.
The question then arising is what is the MAP estimator, or more generally the density, for trajectories of an SDE?
To put this question more precisely, let $I = [0, T]$ be an interval of the real line, and consider the SDE
\beq{equ:continuousdynamics}
\dot{X}_t = f(X_t) + \rho r_t, \qquad t \in I
\eeq
where $f$ is a vector field on $E$, $\rho > 0$, and $r_t, t \in I$ is white noise with zero mean and unit intensity (i.e.\ the correlation function is $\delta(t - s)$ with $\delta$ the Dirac delta function).
Again, we set $X_0 = \xi$, where $\xi \in E$ is known. 
Whatever the precise interpretation of the SDE~\eqref{equ:continuousdynamics}, the solution is a random continuous function $\{X_t, t \in I\}$, and the density of it at some given reference trajectory $\{z_t, t \in I\}$ is defined as
\beq{equ:density_of_SDE}
p(\{z_t\})
 = \lim_{\epsilon \to 0} 
    \frac{\P(\sup_{t \in I}|X_t - z_t| \leq \epsilon)}{\P(\sup_{t \in I} \rho |W_t| \leq \epsilon)}
\eeq
where $\{W_t, t \in I\}$ is the {\em Wiener} process, which can be seen as the time integral of white noise, that is 
\beqn{equ:wiener_I}
W_t = \int_0^t r_s \dd s.
\eeq
We will learn more about the Wiener process later. 
Normalisation with the Wiener process in the Definition~\eqref{equ:density_of_SDE} of the density will turn out to be convenient. 
It is worth stressing that the density in Definition~\eqref{equ:density_of_SDE} is a special case of the Definition~\eqref{equ:densitydef} if we use the norm $\|z\| := \sup_{t \in I} |z_t|$ for trajectories over $I$.
We also note that the density is zero for trajectories which do not start at the initial condition $z_0 = \xi$.
For later use, we introduce the $\epsilon$--{\em weight} 
\beqn{equ:eps_weight}
\alpha(\epsilon, \{z_t\})
 = \frac{\P(\sup_{t \in I}|X_t - z_t| \leq \epsilon)}{\P(\sup_{t \in I} \rho |W_t| \leq \epsilon)}
\eeq
of a trajectory $\{z_t, t \in I\}$. 
The $\epsilon$--weight is the probability that the solution $\{X_t, t \in I\}$ of the SDE~\eqref{equ:continuousdynamics} falls entirely into a small strip or ``sausage'' of width $\epsilon$ around $\{z_t, t \in I\}$, relative to the probability that the Wiener process $\{W_t\}$ falls entirely into a ``sausage'' of width $\epsilon/\rho$ around zero.
Figure~\ref{fig:sausages} illustrates the situation.
The density $p$ according to Definition~\eqref{equ:density_of_SDE} is given by $ p(\{z_t\}) = \lim_{\epsilon \to 0} \alpha(\epsilon, \{z_t\})$.
The density $p$ can be written in the form
\beq{equ:energydensity}
p(\{z_t\})
 = \exp(- \mathcal{A}(\{z_t\})),
\eeq
and several publications seem to imply that $\mathcal{A}(\{z_t\})$ should be equal to the {\em energy functional}
\beq{equ:energyfunctional}
\mathcal{A}_E(\{z_t\}) 
 = \frac{1}{2 \rho^2} \int_I (\dot{z}_t - f(z_t))^2 \dd t,
\eeq
or at least that the MAP estimator should be a minimiser of $\mathcal{A}_E$ (sometimes without clear reference to the concept of densities).
In case observations are present, the energy estimator would carry another term pertaining to the observations.
As mentioned in the introduction already, the correct expression for the functional $\mathcal{A}$ in Equation~\eqref{equ:energydensity} is not the energy functional but the {\em Onsager--Machlup} functional
\beq{equ:onsagermachlupfunctional}
\mathcal{A}_{OM}(\{z_t\}) 
 = \mathcal{A}_{E}(\{z_t\}) + \frac{1}{2} \int_I f'(z_t) \dd t.
\eeq
An informal derivation of this expression will be given in Appendix~\ref{sec:correct_functional}.
Note however that for very small noise amplitudes, the energy functional $\mathcal{A}_{E}$ becomes the dominant term in the Onsager--Machlup functional, as this term scales inversely proportional with the noise, while the additional term does not depend on the noise at all. 
This suggests that data assimilation employing the energy functional does have a rigorous interpretation in the small noise limit. 
This is indeed the case, as discussed for instance in~\cite{vandeneijnden_data_assimilation_low_noise_2012}, where the energy functional emerges from a large deviation principle. 
Furthermore, there are clearly other cases where the additional term in Equation~\eqref{equ:onsagermachlupfunctional} does not matter for the purposes of data assimilation, for instance if the dynamics is linear, as then the second term in  Equation~\eqref{equ:onsagermachlupfunctional} is constant.
In higher dimensions, the additional term is the integral over $\mathrm{div}f(z_t)$ (see Section~\ref{sec:auxilliary}) so that for systems with constant divergence, minimising the energy functional gives the same results as minimising the Onsager--Machlup functional.
In the remainder of this section, we will provide evidence that the expression~\eqref{equ:energydensity} with the energy functional is {\em not} the correct density, and discuss possible reasons for this misconception.
We write the SDE~\eqref{equ:continuousdynamics}, somewhat more rigorously, as an integral equation
\beqn{equ:continuousdynamicsII}
X_t = \xi + \int_0^t f(X_s) \dd s + \rho W_t, \qquad t \in I
\eeq
where $W_t, t \in I$ is the standard Wiener process, which as we have seen can heuristically be interpreted as the integral of the white noise process $r_t$.
In fact, from these heuristics, one can derive that the Wiener process ought to have the following properties:
\begin{enumerate}
\item $W_0 = 0$,
\item for $0 \leq t_1 < t_2$ the {\em increment} $W_{t_2} - W_{t_1}$ is a normally distributed random variable with mean zero and covariance $ t_2 - t_1$,
\item increments for nonoverlapping intervals are independent,
\end{enumerate}
It is well known (see for instance~\cite{BRE},~\cite{moerters_brownian_lectures_2011})
that a process $\{W_t, t \in I\}$ with the properties listed above exists and can be realised as a random continuous function of time.
In view of this, the Equation~\eqref{equ:continuousdynamics} is a classical integral equation perturbed by a randomly selected function that is continuous in time.
Discretisation schemes for Equation~\eqref{equ:continuousdynamics} can be derived by observing that
\beq{equ:continuousdynamicsIII}
X_{t_n} = X_{t_{n-1}} + \int_{t_{n-1}}^{t_n} f(X_s) \dd s + \rho \cdot ( W_{t_n} - W_{t_{n-1}})
\eeq
and approximating the integral in an appropriate way. 
For instance, using the approximation
$ \int_{t_{n-1}}^{t_n} f(X_s) \dd s \cong f(X_{t_{n-1}}) (t_{n} - {t_{n-1}})$ 
and assuming for simplicity a constant time step $(t_{n} - {t_{n-1}}) = \Delta$ results in the Euler scheme \citep[also known as the Euler--Maruyama scheme,][]{milstein_sde_numerics_1995}
\beq{equ:euler}
% s
X^{(\Delta)}_{t_n} = X^{(\Delta)}_{t_{n-1}} + f(X^{(\Delta)}_{t_{n-1}}) \Delta + \rho \cdot ( W_{t_n} - W_{t_{n-1}}).
\eeq
(The superscript $\Delta$ indicates that this solution is obtained with the Euler scheme and time discretisation $\Delta$).
If we set $R_n = \rho \cdot ( W_{t_n} - W_{t_{n-1}})$, then Equation~\eqref{equ:euler} is precisely in the form of Equation~\eqref{equ:discretedynamics} with $F(x) = x + f(x) \Delta$ and $\gamma = \Delta \rho^2$.
Hence the density~\eqref{equ:apostdensity} for the solution $(X^{(\Delta)}_{t_1}, \ldots, X^{(\Delta)}_{t_N})$ of Equation~\eqref{equ:euler} reads as
\beq{equ:apostdensityeuler}
\begin{split} 
& \frac{p_{X^{(\Delta)}_{t_1}, \ldots, X^{(\Delta)}_{t_N}}(x_1, \ldots, x_N)}%
{p_{R_1, \ldots, R_N}(0, \ldots, 0)}\\
& = \exp \Biggl[  -\frac{\Delta}{2\rho^2}
        \sum_{n = 1}^{N} 
            (\frac{x_n - x_{n-1}}{\Delta} - f(x_{n-1}))^2 \Biggr].
\end{split}
\eeq     
It now seems tempting to take the ``limit'' $\Delta \to 0$ here. 
In fact, assuming that the $x_1, \ldots, x_n$ in Equation~\eqref{equ:apostdensityeuler} are the values of some reference trajectory $\{z_t, t \in I\}$ at the points $t_1, \ldots, t_n$, we would by formally taking this limit indeed obtain Equation~\eqref{equ:energydensity} for the density with the energy functional as in Equation~\eqref{equ:energyfunctional}.
If we retrace the steps in our calculation though, we realise that we have not quite taken them in the order we should according to Definition~\eqref{equ:density_of_SDE} of the density.
To discuss this, we introduce the $\epsilon$--{\em weight} of a trajectory $\{z_t, t \in I\}$, but now with respect to the Euler approximation: 
\[
\alpha_{\Delta}(\epsilon, \{z_t\}) = \frac{\P(\sup_{n} |X_{t_n}^{(\Delta)} - z_{t_n}| \leq \epsilon)}{\P(\sup_{n} \rho |W_{t_n}| \leq \epsilon)}.
\]
What we have done to arrive at the Equations~(\ref{equ:energydensity},\ref{equ:energyfunctional}) for the density is to take the limit $\epsilon \to 0$, then use Equation~\eqref{equ:apostdensity} in the special case of the Euler system~\eqref{equ:euler}, and finally take the limit $\Delta \to 0$.
That is, we have proved
\beq{equ:limitwrongway}
\lim_{\Delta \to 0} \lim_{\epsilon \to 0} \alpha_{\Delta}(\epsilon, \{z_t\})
 = \exp (- \mathcal{A}_E(\{z_t\})).
\eeq
However, Definition~\eqref{equ:density_of_SDE} basically requires to take these limits the other way round:
\beq{equ:limitrightway}
p(\{z_t, t \in I\}) = \lim_{\epsilon \to 0}\lim_{\Delta \to 0} \alpha_{\Delta}(\epsilon, \{z_t\}).
\eeq
A simple example (following~\cite{abreu_map_discretisation_2014}) will show that interchanging these two limits will, in general, give different results.
It is evident that the density should be independent of what scheme we use to approximate solutions of SDE's, and the Euler scheme is not the only scheme.
To arrive at another scheme for numerically solving SDE's, we consider other approximations of the integral in Equation~\eqref{equ:continuousdynamicsIII}, for instance
\[
\int_{t_{n-1}}^{t_{n}} f(X_s) \dd s \cong (\lambda f(X_{t_{n-1}}) + (1 - \lambda) f(X_{t_{n}})) \Delta
\]
for some $\lambda \in [0, 1]$, leading to the implicit scheme
\beq{equ:trapez}
X^{(\Delta)}_{t_n} = X^{(\Delta)}_{t_{n-1}} + (\lambda f(X^{(\Delta)}_{t_{n-1}}) + (1 - \lambda) f(X^{(\Delta)}_{t_{n}})) \Delta + \rho \cdot ( W_{t_n} - W_{t_{n-1}}).
\eeq
This is an equally valid approximation scheme for SDE's, see for instance~\cite{kloeden_numerical_sde_1992}, Chapter~12.
Note however that $X^{(\Delta)}_{t_n}$ is now a nonlinear function of the noise $( W_{t_n} - W_{t_{n-1}})$.
Using the same logic as before (see Appendix~\ref{sec:apostdensityII}) one arrives at the conclusion that the functional $\mathcal{A}$ in Equation~\eqref{equ:energydensity} of the density should be
\beq{equ:apostdensityII}
\mathcal{A}_{\lambda} 
 = \exp \Biggl[  -\frac{1}{2\rho^2}
    \int_I
        (\dot{z}_{t} - f(z_{t}))^2 \; \dd t
    - (1 - \lambda) \int_I
            f'(z_{t}) \; \dd t \Biggr].
\eeq     
So not only does another term $- (1 - \lambda) \int_I f'(z_{t}) \; \dd t$ appear in the exponent, but we can generate an entire spectrum of candidate functionals by varying $\lambda$.
This result evidently draws the entire methodology into question. 
We note that $\lambda = 1/2$ gives the Onsager--Machlup functional, that is, $\mathcal{A}_{1/2} = \mathcal{A}_{OM}$.
This however does not prove that $\mathcal{A}_{OM}$ is indeed the correct functional. 
So far, we do not have any reason to believe that $\lambda = 1/2$ is in any way special.
% 
% 
%%%%%%%%%%%%%%%%%% 
% 
\section{Numerical experiment}
\label{sec:numerical_experiment}
\begin{figure}
\begin{center}
\includegraphics{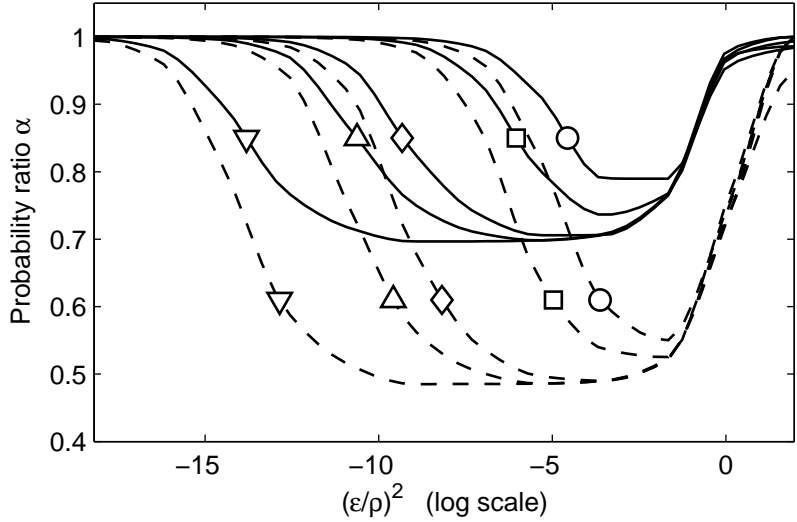}
\caption{\label{fig:exp_density_ratio} The $\epsilon$--weight $\alpha(\epsilon, \Delta)$ as a function of $\epsilon$ for several values of $\Delta$. 
The abscissa shows $\log((\frac{\epsilon}{\rho})^2)$. 
(See text for the reason for this scaling.)
The values for $\log(\Delta)$ are \mbox{-13.7}~($\bigtriangledown$), \mbox{-10.5}~($\bigtriangleup$), \mbox{-9.1}~($\lozenge$), \mbox{-5.9}~($\square$), \mbox{-4.5}~($\bigcirc$).
The solid lines represent results for a shorter time window $T = 0.2$, while the 
dashed lines represent results for $T = 0.4$.}
\end{center}
\end{figure}

It was already mentioned in the last section (and will be discussed further in the Appendix) that $\mathcal{A}_{OM}$ is the appropriate density functional for paths of a stochastic differential equation.
In particular, this implies that the minimiser of $\mathcal{A}_{OM}$ can be interpreted as the MAP estimator or ``most probable'' path of the stochastic differential equation.
We have also considered discrete time approximations to the stochastic differential equations, for instance the Euler scheme, and it emerged that the densities derived from these approximations do not, in general, agree with the Onsager--Machlup functional even approximately. 
This raises questions as to what the right functional should be {\em in practice}, since apart from the rare situation where explicit solutions are available, stochastic differential equations inevitably have to be approximated by numerical schemes which are discrete in time.
But suppose we approximate a stochastic differential equation of the form~\eqref{equ:continuousdynamics} with the Euler scheme~\eqref{equ:euler}.
We know that in this situation, Equation~\eqref{equ:apostdensityeuler} is the {\em correct} density of solutions, so what is the link between solutions of the Euler scheme and the functional $\mathcal{A}_{OM}$, and why should we care about it?
We will examine the situation with a numerical example.
We consider a stochastic differential equation of the form~\eqref{equ:continuousdynamics} with approximation by the Euler scheme (Equ.~\ref{equ:euler}).
Here $f(x) = \frac{2}{\pi}\arctan(a x)$, with $a = 6$ and $\rho = 0.3$.
All solutions start from the fixed initial condition $\xi = 0$.
\begin{figure}
\begin{center}
\includegraphics{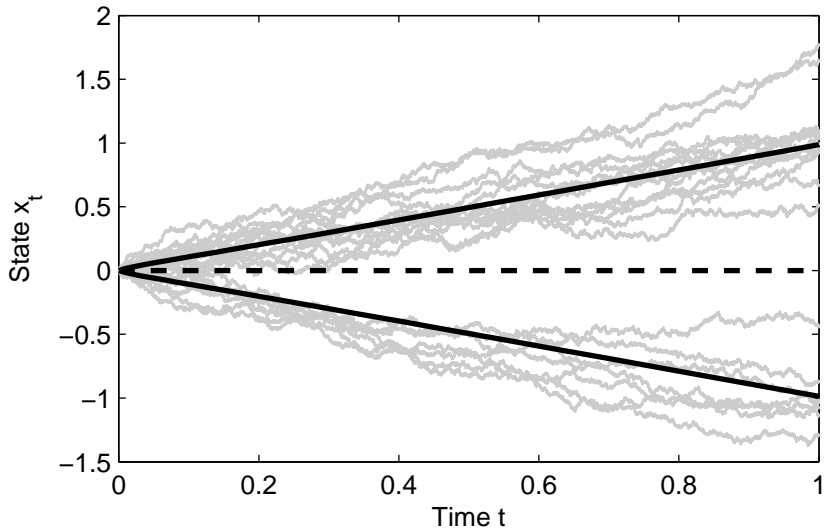}
\caption{\label{fig:examplesimulation} Around~20 simulations of the SDE~\eqref{equ:continuousdynamics} with $f(x) = \frac{2}{\pi}\arctan(a x)$ and $a = 6$ and $\rho = 0.3$ are shown in grey, obtained with an Euler scheme with $\Delta = 1.14\cdot 10^{-4}$. 
The two solid lines represent the most probable trajectories according to the Onsager--Machlup functional, and the dashed line represents the most probable trajectory according to the energy functional. 
It is evident that simulations are more likely to accumulate around the former.}
\end{center}
\end{figure}
Figure~\ref{fig:examplesimulation} shows 20~independent approximate solutions of Equation~\eqref{equ:continuousdynamics}; ``approximate'' because these are solutions of the Euler scheme~\eqref{equ:euler}.
The density of these solutions is given by Equation~\eqref{equ:apostdensityeuler}, and according to this expression the most probable solution is equal to zero for all times. 
The picture though we see in Figure~\ref{fig:examplesimulation} seems to contradict this.
It is evident that very few solutions seem to be concentrating around the abscissa.
This is easy to understand qualitatively. 
For small times, the variability of the solution grows exponentially as the origin is an unstable fixed point for this dynamics. 
Sooner or later, the solution will enter regions where the arctan is flat and the drift is essentially either $+1$ or $-1$. 
The solution might from time to time transit between these two regimes, but these transits become progressively rarer until it behaves essentially like a random walk with constant drift. 
The solid lines in Figure~\ref{fig:examplesimulation} represent the optimal paths of the Onsager--Machlup functional $\mathcal{A}_{OM}$.
These have been calculated numerically by solving the Euler Lagrange equations associated with the Onsager--Machlup functional $\mathcal{A}_{OM}$ (the functional displays a symmetry whence there are two solutions symmetric about the abscissa).
These solutions seem to capture much better the ``big picture'', indicating where solutions of our simulations tend to be.
So it seems that the Onsager--Machlup functional provides a better description of the density, even though the solutions have been obtained with a discrete time system and thus strictly speaking Equation~\eqref{equ:apostdensityeuler} provides the correct density.
To resolve this apparent paradox, we remember that the density at some reference path $\{z_t\}$ is the probability that the solution of our dynamics lies in a thin sausage of width $\epsilon$ around that reference path, relative to the probability that the driving Wiener process lies in a thin sausage of width $\epsilon$ around zero.
These probabilities, or rather the $\epsilon$--weight $\alpha_\Delta(\epsilon, \{z_t\})$ can be estimated using a Monte Carlo approach in order to study the dependence on $\epsilon$ and $\Delta$.  
For simplicity, the reference path was taken to be zero.
% %%
Note that this is the most probable path according to $\mathcal{A}_{E}$.
In Figure~\ref{fig:exp_density_ratio}, $\alpha_\Delta(\epsilon, \{z_t\})$ is shown as a function of $\epsilon$ (on the abscissa), with different curves (different marker symbols) corresponding to different values of $\Delta$ (curves corresponding to smaller values of $\Delta$ tend to be more to the left in the plot).
Two time windows of different length were considered; the solid lines correspond to $T = 0.2$, while the dashed lines correspond to an experiment with $T = 0.4$.
The discussion in Section~\ref{sec:map_estimators_sde} revealed that taking the limits $\lim_{\epsilon \to 0}$ and $\lim_{\Delta \to 0}$ of $\alpha_\Delta(\epsilon, \{z_t\})$ in different order gives different results, see Equations~(\ref{equ:limitwrongway},\ref{equ:limitrightway}).
Along the particular path considered here, $\mathcal{A}_{E} = 0$ (independent of the value of $\Delta$), meaning that % 
\[
\lim_{\Delta \to 0} \lim_{\epsilon \to 0} \alpha_\Delta(\epsilon, \{z_t\}) = \exp(-\mathcal{A}_{E}) = 1,
\]
while interchanging these limits gives the values $\exp(-\mathcal{A}_{OM}) = 0.68$ for $T = 0.2$ and $\exp(-\mathcal{A}_{OM}) = 0.47$ for $T = 0.4$ (obtained by simply evaluating $\mathcal{A}_{OM}$ along our reference path).
% 
%%%
The fact that interchanging the limits gives different results manifests itself in the plot in Figure~\ref{fig:exp_density_ratio} which shows an interesting crossover behaviour. 
With $\epsilon$ decreasing, $\alpha$ first approaches the value given by the Onsager--Machlup functional. 
If $\epsilon$ reaches a sufficiently small value though (depending on $\Delta$), the curves start to diverge from this value and approach one, consistent with the energy functional.
The smaller $\Delta$, the longer $\alpha$ stays close to the Onsager--Machlup value for decreasing $\epsilon$, or in other words, for smaller $\Delta$ a smaller $\epsilon$ has to be chosen in order for $\exp(-\mathcal{A}_E)$ to become a relevant approximation for $\alpha$.
For a rough estimate on how small $\epsilon$ has to be in order for the crossover to take place, we observe that for a reference path $z$,
\beq{equ:6.10}
X^{(\Delta)}_{t_{n+1}} - z_{t_{n+1}}
 = X^{(\Delta)}_{t_n} - z_{t_{n}} 
 + \left(f(X^{(\Delta)}_{t_n}) - \frac{z_{t_{n+1}} - z_{t_{n}}}{\Delta}\right)\Delta + \rho (W_{t_{n+1}} - W_{t_n}).
\eeq
Hence for fixed $\Delta$, the increments of $X^{(\Delta)}_{t_{n}} - z_{t_{n}}$ in Equation~\eqref{equ:6.10} have a characteristic size (at time $t_n$), namely
\beqn{equ:6.20}
|X^{(\Delta)}_{t_{n+1}} - z_{t_{n+1}}
    - (X^{(\Delta)}_{t_n} - z_{t_{n}})| 
\cong
|f(z_{t_n}) - \dot{z}_{t_{n}}|\Delta + \rho \sqrt{\Delta},
\eeq
It seems plausible that $\alpha$ starts to approach the energy functional as soon as $\epsilon$ becomes smaller than the typical increment of $X^{(\Delta)}_{t_{n}} - z_{t_{n}}$, which means 
\[
\epsilon \cong |f(z_{t_n}) - \dot{z}_{t_{n}}|\Delta + \rho \sqrt{\Delta},
\]
which is just $\epsilon \cong \rho \sqrt{\Delta}$ in our case, or $\log((\frac{\epsilon}{\rho})^2) = \log(\Delta)$. 
For the experiments shown in Figure~\ref{fig:exp_density_ratio}, we used the following values of $\log(\Delta)$: \mbox{-13.7}~($\bigtriangledown$), \mbox{-10.5}~($\bigtriangleup$), \mbox{-9.1}~($\lozenge$), \mbox{-5.9}~($\square$), \mbox{-4.5}~($\bigcirc$).
This appears to be roughly consistent with the values of of $\log((\frac{\epsilon}{\rho})^2)$ at which the crossover takes place.
% 
% %%%%%%%%%%%%%%%%%%%%%%%%%
% 
\section{The Onsager--Machlup functional in higher dimensions and for multiplicative noise}
\label{sec:auxilliary}
In this section we will provide additional (and well known) results regarding the Onsager--Machlup functional in higher dimensions and with multiplicative noise.
We will see that in the case of multiplicative noise, further terms appear in the Onsager--Machlup functional; the effect of these terms in data assimilation applications remains to be investigated.
We consider a general SDE
\beq{equ:7.10}
\dot{X}_t = f(X_t) + \rho(X_t) \cdot r_t, \qquad t \in [0, T]  
\eeq
where the state space $E$ is the $d$--dimensional Euclidean space, $f$ is a vector field on $E$ and $\rho$ is a state~dependent $d$--by--$d$ matrix.
For SDE's with multiplicative noise as in Equation~\eqref{equ:7.10}, different mathematical interpretations are possible, most prominently the It\^{o} and the Stratonovi\v{c} interpretation~\citep[see e.g.][]{oksendal_SDE_1998,ikeda_sde_diffusion_1989}.
We will interprete the SDE~\eqref{equ:7.10} in the sense of Stratonovi\v{c}; and  It\^{o}~equation can always be converted to a  Stratonovi\v{c}~equation.
The expression for the Onsager--Machlup functional given in Equation~\eqref{equ:7.30} below is valid if the noise is nondegenerate, that is $\rho(x) \rho^T(x) \geq \alpha \cf$ for some $\alpha > 0$.
In this situation, the matrix $g(x) = (\rho(x) \rho^T(x))^{-1}$ defines a Riemannian metric. 
For any vector field $f$, the divergence $\mathrm{div}f$ will be understood with respect to this metric, that is
\beqn{equ:7.20}
\mathrm{div}f = \frac{1}{ \sqrt{|g|}} \sum_{k = 1}^d \partial_k ( \sqrt{|g|} f^{(k)})
\eeq
Further, let $R(x)$ be the scalar (Ricci) curvature and $m(x, y)$ the (geodesic) distance between points $x, y \in E$.
These concepts are defined with respect to the metric $g$ as well~\citep[see][for an introduction to Riemannian geometry]{gallot_riemannian_geometry_2004}.
Then the Onsager--Machlup functional is defined as
\beq{equ:7.25}
\exp (- \OM (\{z_t\})) 
 = \lim_{\epsilon \to 0}
    \frac{\P(\sup_{t \in I} m(X_t, z_t) \leq \epsilon)}{\P(\sup_{t \in I} |W_t| \leq \epsilon)},
\eeq
and as is proved for instance in~\citet{ikeda_sde_diffusion_1989,zeitouni_maximum_1987}, it has the expression
\beq{equ:7.30}
\begin{split}
\OM (\{z_t\})
 & = \frac{1}{2} \int_I (\dot{z}_t - f(z_t))^T \! g(z_t) (\dot{z}_t - f(z_t)) \; \dd t \\
 & \qquad + \frac{1}{2} \int_I \mathrm{div}f(z_t) \dd t
    - \frac{1}{12} \int_I R(z_t) \; \dd t.
\end{split}
\eeq
As was already discussed in Section~\ref{sec:numerical_experiment} (in the context of a one--dimensional example), the effect of the second term (containing $\mathrm{div}f$) is to discourage the most probable path from staying in regions where the dynamics is unstable, as this causes strong amplification of the noise and thus typical solutions of the SDE quickly escape from such regions.
The effect of the second term involving the Ricci curvature is not so clear at this point and is subject to future investigation.
In the remainder of this section we discuss what terms need adding to the Onsager--Machlup functional if observations are present. 
Let the observations be a discrete time series $\{Y_n, n = 1, \ldots, N\}$. 
The Onsager--Machlup functional is now defined as 
\beqn{equ:7.40}
\exp (- \OMF (\{z_t\}, \{Y_n\})) 
 = \lim_{\epsilon \to 0}
    \frac{\P(\sup_{t \in I} m(X_t, z_t) \leq \epsilon|\{Y_n, n = 1, \ldots, N\})}{\P(\sup_{t \in I} |W_t| \leq \epsilon)}.
\eeq
(We will use the notation $\OMF$ to designate the Onsager--Machlup functional with observations; $\OM$ still defined as in Eq.~\ref{equ:7.25}.)
A commonly made assumption is that the observations are conditionally independent given the underlying trajectory $\{X_t, t \in [0, T]\}$, and that the distribution of $Y_n$ depends on $X_{t_n}$ only for $n = 1, \ldots, N$ and a series of sampling times $t_1, \ldots, t_N$.
Let $q_n(y, x)$ be the density of $Y_n$ given $X_{t_n}$.
In this case, the full Onsager--Machlup functional reads as 
\beqn{equ:7.40}
\OMF(\{z_t\}; \{Y_n\})
%  = \OM(\{z_t\}) + \OMO (\{z_t\}; \{Y_n\})
 = \OM(\{z_t\}) - \sum_{n = 1}^N \log(q_n(Y_n, z_{t_n}))
\eeq
If for instance $Y_n$ given $X_{t_n}$ is Gaussian with mean $h(X_{t_n})$ and variance $\gamma$ (where $h$ and $\gamma$ are often called the observation function and observation error covariance, respectively), then the additional term in the Onsager--Machlup functional reads as 
\beqn{equ:7.40}
- \sum_{n = 1}^N \log(q_n(Y_n, z_{t_n}))
 = \frac{1}{2} \sum_{n = 1}^N (y_n - h(z_{t_n}))^T\gamma^{-1}(y_n - h(z_{t_n})).
\eeq
% 
% 
% %%%%%%%%%%%%%%%%%%%%%%%%%
% 
\section{Conclusions for discrete time simulations and data assimilation}
\label{sec:conclusions}
When modelling a dynamical process with a stochastic differential equation, then any practical implementation will use a discrete time approximation of one form or another. 
If (as part of a data assimilation experiment for instance) one is interested in a most probable path of that dynamical process, then our considerations imply that the appropriate functional is the Onsager--Machlup functional (or a discrete time approximation thereof), even though the density of discrete time approximations might differ from the Onsager--Machlup functional. 
The Onsager--Machlup functional provides results which are robust with respect to the particular approximation scheme, and in particular with respect to the chosen time discretisation, which does not have any intrinsic meaning in terms of the problem specification.
More specifically, the Onsager--Machlup functional gives approximately the $\epsilon$-weight of a reference path, that is the probability that the solutions of the stochastic differential equation stay in an $\epsilon$ sausage around the reference path, and a discrete time approximation of the SDE will assign approximately the same $\epsilon$-weight to that path, {\em unless} $\epsilon$ reaches the scale of typical increments in that approximation.
In other words, the Onsager--Machlup functional provides an approximation to the $\epsilon$-weight of a path with respect to the stochastic differential equation {\em and} approximations thereof, save approximations that employ increments which are typically larger than $\epsilon$.  
Such approximations do not appropriately represent the fast fluctuations of the Wiener process that are still relevant for the dynamics, even when the amplitude of Wiener process is constrained to be small.
For these reasons, most probable paths should be determined using the Onsager--Machlup functional, since such paths carry the largest possible $\epsilon$-weight, no matter if this weight is calculated from the stochastic differential equation or any reasonable approximation, as long as that approximation uses increments which are smaller than $\epsilon$.  
Paths which are minimisers of the energy functional or any other functional do not possess this universality property.
The implication for data assimilation is that minimising paths of the Onsager--Machlup functional are more typical for the dynamics and in fact carry a rigorous interpretation as MAP estimators, different from maximum energy paths which do not.
These arguments do not apply though if the process under consideration is intrinsically discrete in time. 
In this situation, it does not make sense to consider the limit $\Delta \to 0$ which brings about the extra term in the Onsager--Machlup functional. 
Systems like this might appear in the context of seasonal or diurnal cycles, or more generally systems with an internal clocking mechanism.
\appendix
% 
% %%%%%%%%%%%%%%%%%%%%%%%%%%%%%%%%%%%%%%%%%%
% 
\section*{Appendix}
\section{Derivation of the correct functional}
\label{sec:correct_functional}
We will attempt a more careful calculation of the $\epsilon$--weight of a path which will not only allow us to take the limits in the right order and obtain the correct expression for the density, but also to identify the reason why interchanging these limits gives a different result.
We will later restrict our attention to linear dynamics. 
It should be said that for linear dynamics, the additional term in the Onsager--Machlup functional~\eqref{equ:onsagermachlupfunctional} does not depend on the reference trajectory and hence minimising $\mathcal{A}_{OM}$ or $\mathcal{A}_{E}$ gives the same results in this case. 
However, the functionals are still different and only the Onsager--Machlup functional provides the correct density.
First we note the following simple but important fact.
Let $Z^{(1)}, Z^{(2)}$ be random variables with values in $\R^N$ with densities $p_1, p_2$ respectively, and $p_2(z) > 0$ for all $z \in \R^N$.
Further, let $\phi$ be a function on $\R^N$. 
Then the identity
\beqn{equ:5.30}
\E(\phi(Z^{(1)})) = \E(\phi(Z^{(2)}) \frac{p_1(Z^{(2)})}{p_2(Z^{(2)})} )
\eeq
holds, since
\beq{equ:5.40}
\E(\phi(Z^{(1)}))
 = \int_{\R^n}\phi(z) p_1(z) \dd z 
 = \int_{\R^n} \phi(z) \frac{p_1(z)}{p_2(z)} p_2(z) \dd z
 = \E(\phi(Z^{(2)}) \frac{p_1(Z^{(2)})}{p_2(Z^{(2)})} ).
\eeq
On the other hand, note that
\beq{equ:5.45}
\P(\max_k | X_{t_k} - z_{t_k}| \leq \epsilon)
 = \E (H(\frac{\max_k | X_{t_k} - z_{t_k}|}{\epsilon} - 1)),
\eeq
where $H$ is the Heaviside function.
We might use Equation~\eqref{equ:5.40} in~\eqref{equ:5.45} with 
\beqn{equ:5.50}
\begin{split}
\phi(z) & = H(\frac{\max_k |z_k|}{\epsilon} - 1), \\
Z^{(1)} & = (X^{(\Delta)}_{t_1} - z_{t_1}, \ldots, X^{(\Delta)}_{t_N} - z_{t_N}),\\
Z^{(2)} & = (W_{t_1}, \ldots, W_{t_N}),
\end{split}
\eeq
where $(X^{(\Delta)}_{t_1}, \ldots, X^{(\Delta)}_{t_N})$ is a solution to the Euler approximation~\eqref{equ:euler}.
Note that $(X^{(\Delta)}_{t_1} - z_{t_1}, \ldots, X^{(\Delta)}_{t_N} - z_{t_N})$ is then a solution of the system~\eqref{equ:6.10}.
We therefore obtain
\beq{equ:5.60}
\P(\max_k|X_{t_k} - z_{t_k}| \leq \epsilon) 
= \E\Big[ H(\frac{\max_k |W_{t_k}|}{\epsilon} - 1)
    \exp (  A + B + C )
    \Big]
\eeq
with
\beq{equ:5.70}
\begin{split}
A & = -\frac{\Delta}{2\rho^2}
        \sum_{n = 1}^{N} 
            (\frac{z_{t_n} - z_{t_{n-1}}}{\Delta} - f(z_{t_{n-1}} + \rho W_{t_{n-1}}))^2\\
B & = - \frac{1}{\rho} \sum_{n = 1}^{N}
            (\frac{z_{t_n} - z_{t_{n-1}}}{\Delta})
            (W_{t_n} - W_{t_{n-1}})\\
C & = \frac{1}{\rho} \sum_{n = 1}^{N} 
            (f(z_{t_{n-1}} + \rho W_{t_{n-1}}))
            (W_{t_n} - W_{t_{n-1}})
\end{split}
\eeq
In terms of the limits $\Delta \to 0$ and $\epsilon \to 0$, the first two terms $A$ and $B$ will converge to 
\beq{equ:5.80}
\lim_{\epsilon \to 0} \lim_{\Delta \to 0} A = -\frac{1}{2\rho^2}
        \int_{0}^{T} 
            (\dot{z_t} - f(z_t))^2 \dd t
\eeq
and zero, respectively, no matter in which order the limits are taken. 
The third term however shows different behaviour depending on whether $\Delta \to 0$ or $\epsilon \to 0$ first. 
If we take $\Delta \to 0$ first, it can be shown that a well defined random variable obtains\footnote{The limit is in fact in the $L_2$~sense.} which can be written as an Ito integral
\beqn{equ:5.90}
\lim_{\Delta \to 0} C 
= \frac{1}{\rho } \int_{0}^{T} f(z_{t} + \rho W_{t}) \dd W_{t}.
\eeq
We do not expect the reader to be familiar with the theory of Ito integrals -- relevant here is that the limit of this expression for $\epsilon \to 0$ will {\em not} be zero but
\beq{equ:5.100}
\lim_{\epsilon \to 0}  \lim_{\Delta \to 0} C 
= -\frac{1}{2} \int_{0}^{T} f'(z_{t}) \dd t.
\eeq
A demonstration of this fact\footnote{%
Strictly speaking this ``fact'' is only correct in a much weaker sense but still sufficient to derive the Onsager--Machlup functional; 
The correct statement is that
\[
\E \Bigl[
    \exp \left( 
        \int_{0}^{T} f(z_{t} + W_{t}) \dd W_{t} 
        + \frac{1}{2} \int_{0}^{T} f'(z_{t}) \dd t
    \right)  
    \Big\vert \sup_t |W_t | \leq \epsilon 
\Bigr] 
\to 1
\]
for $\epsilon \to 0$, see~\cite{ikeda_sde_diffusion_1989}.
}
for the case where $f$ is linear is given here for illustration.
If $f(x) = a x$ for some $a \in \R$, then
\beq{equ:5.110}
\begin{split}
C & = \frac{a}{\rho} \sum_{n = 1}^{N} 
            (z_{t_{n-1}} + \rho W_{t_{n-1}})
            (W_{t_n} - W_{t_{n-1}})\\
 & = \frac{a}{\rho} \sum_{n = 1}^{N} 
            z_{t_{n-1}}
            (W_{t_n} - W_{t_{n-1}})
 + a \sum_{n = 1}^{N} 
            W_{t_{n-1}}
            (W_{t_n} - W_{t_{n-1}})\\
 & = \frac{a}{\rho} C_1 + a C_2.
\end{split}
\eeq
It is easy to see that $C_1 \to 0$ if $\Delta \to 0$ and $\epsilon \to 0$, no matter in which order these limits are taken. 
After some algebra, we can write $C_2$ as 
\beqn{equ:5.120}
\begin{split}
C_2 & = \sum_{n = 1}^{N} 
            W_{t_{n-1}}
            (W_{t_n} - W_{t_{n-1}})\\
 & = \frac{1}{2} W_{T}^2 
        - \frac{1}{2} \sum_{n = 1}^{N} (W_{t_n} - W_{t_{n-1}})^2
\end{split}
\eeq
Considering the mean and the variance of the second term, we obtain $\frac{1}{2} T$ and $\frac{1}{2} T\Delta$, respectively, implying that (at least in a mean square sense) the second term converges to its mean $\frac{1}{2} T$ if $\Delta \to 0$.
Hence
\beq{equ:5.130}
\lim_{\Delta \to 0} C_2 
 = \frac{1}{2} W_{T}^2
 - \frac{1}{2} T
\eeq
Therefore, taking the limits $\Delta \to 0$ and then $\epsilon \to 0$ in Equation~\eqref{equ:5.110} and using Equation~\eqref{equ:5.130} we obtain
\beqn{equ:5.140}
\lim_{\epsilon \to 0} \lim_{\Delta \to 0}C
 = -\frac{a}{2} T
\eeq
which is the same as Equation~\eqref{equ:5.100} for this special case.
Using Equation~\eqref{equ:5.100} and the expression in Equation~\eqref{equ:5.80} in~\eqref{equ:5.60} we obtain that for small $\epsilon$ 
\beqn{equ:5.150}
\begin{split}
\P(\sup_t |X_{t} - z_{t}| \leq \epsilon ) 
& \cong  \E( H(\frac{\sup_t|W_{t}|}{\epsilon} - 1) \\
& \quad \cdot \; \exp 
    \Big[
    -\frac{1}{2\rho^2} \int_{0}^{T} 
            (\dot{z_t} - f(z_t))^2 \dd t
    -\frac{1}{2} \int_{0}^{T} 
    f'(z_{t}) \dd t.
    \Big]
\end{split}
\eeq
so that we can conclude
\beqn{equ:5.160}
\begin{split}
& \lim_{\epsilon \to 0}\frac{\P(\sup_t |X_{t} - z_{t}| \leq \epsilon)}{%
    \P( \sup_t|W_{t}| \leq \epsilon)}\\
& = \exp 
    \Big[
    -\frac{1}{2\rho^2} \int_{0}^{T} 
            (\dot{z_t} - f(z_t))^2 \dd t
    -\frac{1}{2} \int_{0}^{T} 
    f'(z_{t}) \dd t.
    \Big]\\
& = \exp(-\mathcal{A}_{OM}).
\end{split}
\eeq
Note that if we used Equations~(\ref{equ:5.60},\ref{equ:5.70}) as a starting point but subsequently took the limits in the wrong order, that is, first $\epsilon \to 0$ and then $\Delta \to 0$, we would have $B, C \to 0$, so we would obtain the energy estimator $\mathcal{A}_{E}$.
As a final remark, by looking back at the calculations the reader will see that the only term that does not permit interchange of the limits is a second order or ``quadratic'' term
$ \sum_{n = 1}^N (W_{t_n} - W_{t_{n-1}})^2$
which would vanish with $\Delta \to 0$ if $W$ were a differentiable function but converges to $T$ in case of the Wiener process.
Roughly speaking, this is because $W_{t_n} - W_{t_{n-1}}$ is of order $\sqrt{\Delta}$, which more generally gives rise to the extra terms in the Ito calculus.
\section{Derivation of Equation~\eqref{equ:apostdensityII}}
\label{sec:apostdensityII}
In this section, we will derive the Equation~\eqref{equ:apostdensityII}, that is, we follow same steps as for the Euler scheme and take the limits as in Equation~\eqref{equ:limitwrongway}, but starting with the implicit scheme~\eqref{equ:trapez} instead of the Euler scheme.
If we set $R_n = \rho (W_{t_n} - W_{t_{n-1}})$, then  the implicit scheme~\eqref{equ:trapez} can be written in the form
\beqn{equ:b.20}
X_{t_{n}} = X_{t_{n-1}} + F_1(X_{t_{n}}) + F_2(X_{t_{n-1}}) + R_n
\eeq
which can be expressed as $(R_1, \ldots, R_N) = \Psi(X_{t_1}, \ldots, X_{t_N})$ with 
\beqn{equ:b.30}
\Psi_n(x_1, \ldots, x_N) = x_{n} - x_{n-1} - F_1 (x_{n}) - F_2(x_{n-1}) \qquad \text{for $n = 1, \ldots, N$}.
\eeq
According to basic probability calculus, we have for the densities
\beq{equ:b.40}
p_{X_{t_1}, \ldots, X_{t_N}}(x_{1}, \ldots, x_{N})
 = p_{R_1, \ldots, R_N}(\Psi(x_{1}, \ldots, x_{N})) \cdot 
	\left| \frac{\partial \Psi}{\partial x}(x_{1}, \ldots, x_{N}) \right|
\eeq
Since $\frac{\partial \Psi_k}{\partial x_l} = 0$ for $k < l$, the Jacobi~matrix of $\Psi$ is lower left triangular and hence
\beqn{equ:b.50}
\begin{split}
\left| \frac{\partial \Psi}{\partial x} \right|(x_{1}, \ldots, x_{N})
 & = \prod_{n = 1}^N \frac{\partial \Psi_k}{\partial x_k}(x_{1}, \ldots, x_{N}) \\
 & = \prod_{n = 1}^N 1 - F_1'(x_k) \\
 & = \prod_{n = 1}^N 1 - (1 - \lambda) \Delta f'(x_k) \\
 & = \exp \left( \sum_{n = 1}^N \log(1 - (1 - \lambda) \Delta f'(x_k)) \right).
\end{split}
\eeq
We evaluate this expression with $x_k = z_{t_k}$ for $k = 1, \ldots, N$ where $\{z_t\}$ is some trajectory on the interval $I = [0, T]$ and $N = T/\Delta$. 
Since $\log(1 + w) \cong w$ for small $w$, we can write the exponent approximately as 
\beqn{equ:b.60}
\sum_{n = 1}^N \log(1 - (1 - \lambda) \Delta f'(z_{t_n}))
\cong - (1 - \lambda) \Delta  \sum_{n = 1}^N  f'(z_{t_n})
\eeq
which is a Riemann sum converging to $-(1 - \lambda)\int_I f'(z_t) \; \dd t$.
The first factor in Equation~\eqref{equ:b.40}, after normalisation and when evaluated along a trajectory, reads as
\beqn{equ:b.60}
\begin{split}
& \frac{p_{R_1, \ldots, R_N}(\Psi(z_{t_1}, \ldots, z_{t_N}))}{p_{R_1, \ldots, R_N}(0, \ldots, 0) }\\
&  = \exp \left( -\frac{\Delta}{2 \rho^2}
	\sum_{n = 1}^N \left( 
		\frac{z_{t_n} - z_{t_{n-1}}}{\Delta} - (1 - \lambda) f(z_{t_n}) - \lambda f(z_{t_{n-1}})
	\right)^2
\right).
\end{split}
\eeq
Again, the exponent is a Riemann sum which converges to $-\frac{1}{2 \rho^2}\int_I (\dot{z}_t - f(z_t))^2 \dd t$ for $\Delta \to 0$.
In summary, we get Equation~\eqref{equ:apostdensityII}.
% 

%
% \bibliographystyle{plainnat}
% \bibliography{../../../TeX/Literatur}
% 
% 
\end{document}